\documentclass[year=24]{fmcad}
\usepackage[T1]{fontenc}
\usepackage{graphicx}
\usepackage[outdir=./]{epstopdf}
\usepackage{hyperref}
\usepackage{listings}
\usepackage{amsmath}
\usepackage{amssymb}
\usepackage{bussproofs}
\usepackage{qtree}
\usepackage{varwidth}
\usepackage{multirow}
\usepackage[font=scriptsize]{subcaption}
\usepackage[justification=Centering]{caption}
\usepackage[nocompress]{cite}

\newcommand{\proves}{\vdash}
\newenvironment{bprooftree}
  {\leavevmode\hbox\bgroup}
  {\DisplayProof\egroup}
\newcommand{\fun}[2]{{\tt\bf fun}(#1, #2)}

\begin{document}

\title{Modernizing SMT-Based Type Error Localization}
\author{
\authorblockN{Max Kopinsky \orcid{0009-0005-6835-511X}}
\authorblockA{McGill University\\
Montréal, Quebec\\
max.kopinsky@mail.mcgill.ca}
\and
\authorblockN{Brigitte Pientka \orcid{0000-0002-2549-4276}}
\authorblockA{McGill University\\
Montréal, Quebec\\
bpientka@cs.mcgill.ca}
\and
\authorblockN{Xujie Si \orcid{0000-0002-3739-2269}}
\authorblockA{University of Toronto\\
Toronto, Ontario\\
CIFAR AI Research Chair\\
six@cs.toronto.edu}}

\maketitle              % typeset the header of the contribution
\begin{abstract}
Traditional implementations of strongly-typed functional programming languages often miss the root cause of type errors. As a consequence, type error messages are often misleading and confusing - particularly for students learning such a language.
We describe Tyro, a type error localization tool which determines the optimal source of an error for ill-typed programs following
fundamental ideas by Pavlinovic et al. : we first translate typing constraints into SMT (Satisfiability Modulo Theories) using an intermediate representation which is more readable than the actual SMT encoding; during this phase we apply a new encoding for polymorphic types.
Second, we translate our intermediate representation into an actual SMT encoding and 
take advantage of recent advancements in off-the-shelf SMT solvers to effectively find optimal error sources for ill-typed programs.
Our design maintains the separation of heuristic and search also present in prior and similar work. In addition, our architecture design increases modularity, re-usability, and trust in the overall architecture using an intermediate representation to facilitate the safe generation of the SMT encoding. We believe this design principle will apply to many other tools that leverage SMT solvers.

Our experimental evaluation reinforces that the SMT approach finds accurate error sources using both expert-labeled programs and an automated method for larger-scale analysis. Compared to prior work, Tyro lays the basis for large-scale evaluation of error localization techniques, which can be integrated into programming environments and enable us to understand the impact of precise error messages for students in practice.

% Ensure that author names don't get dashes in bib
% Must appear before any \cites.
\bstctlcite{bstctl:nodash}
%\keywords{MaxSMT \and Type Inference \and Error Diagnostics}

\end{abstract}
\section{Introduction}

%Introduction: 
%% Traditional implementations of strongly-typed functional programming languages often miss the root cause of type errors. As a consequence, type error messages are often misleading and confusing -- particularly for students learning such a language.

%% 2-3 paragraphs : discussing approaches to type error diagnostics / localization  (typical one, machine learning based on E. Seidel et al, Mycroft ....) -- setting the stage

%% OUR work: In contrast to prior work, we purse SMT based approach --> Minerr

%%  Contribution:
%%  1. Tyro, a type error localization tool which determines the optimal source of an error for ill-typed programs following
%%     fundamental ideas by Pavlinovic et al. : we first translate typing constraints into SMT (Satisfiability Modulo Theories) using an intermediate representation which is more readable than the actual SMT encoding; during this phase we apply a new encoding for polymorphic types.

%%  2. Experimental evaluation;
%%     a) Base line: our tool is as good as Minerr 
%%     b) evaluation wrt hand-labelled data 
%%     c) evaluation wrt to student's programs and how they fixed a given error

Many strongly typed programming languages, such as OCaml~\cite{OCaml},
allow programmers to omit type annotations from their code; despite these omissions,
%% the language is able to maintain strong static typing via \emph{type inference},
\emph{type inference} automatically reconstructs the types of all expressions in the program based
on the contexts in which they appear. For well-typed programs,
type inference saves the programmer much time and effort. However, for ill-typed
programs, the situation can be exactly the opposite~\cite{HowStudentsFixedTypeErrors}. Type errors are discovered when the compiler
%% is unable to find a solution to the type inference problem, but figuring out \emph{why}
%% there is no solution is much harder. In particular, it is difficult to discover
finds inconsistencies during type inference, but figuring out \emph{root causes} is much harder. 
The location where compiler fails is usually \emph{not} the place to fix the reported type errors. 
As a result, type errors are often misleading or confusing. Such errors increase debugging time for programmers.
In the case of novices, such errors discourage them from learning the language at all~\cite{SearchingForErrors}.
Even tools designed to assist novices, such as Helium~\cite{Helium}, frequently produce such misleading errors.

The importance, and difficulty, of finding accurate causes of type errors 
(``localization'') has a long-studied history. A system for recording ``reasons'' that
may explain type mismatches was implemented in Wand's SPS~\cite{SPS} in 1986~\cite{FindingTheSource}. % This method does not pick a location; contrast with Flow in related work.
Improvements to Wand's method include the recent HM$^\ell$, which turns the problem
of explaining the ``reasons'' into a data flow problem \cite{GettingIntoTheFlow}.
Other recent approaches use machine learning techniques to localize errors~\cite{LearningToBlame,NoviceTypeErrorDiagnosis} but without any formal guarantees.

There is also a class of techniques based on heuristic search. Type inference is naturally
expressed as a constraint-solving problem~\cite{ConstraintForm,EssenceOfML,EfficientGeneralization}, even for more complex type systems, e.g.~\cite{TypeInferenceConstrained}. %TypeInferenceForArbitraryRank?
By heuristically attributing weights to each constraint, techniques for constrained
optimization can be applied. Such techniques can involve custom frameworks and solvers,
as in Mycroft~\cite{Mycroft}; or more generalized tools such as SMT solvers.

Our work builds on prior work using SMT solvers. 
% SMT, or ``Satisfiability Modulo Theories,''
% is a general approach to solving constraint problems, and has a standard lisp-like interface for solvers~\cite{SMTLIB2-6}.
Cutting-edge SMT solvers, such as Z3~\cite{Z3}, are being actively developed and steadily improved.
These improvements cut down on memory usage and runtime, enabling SMT solvers to handle
increasingly large problem instances.
Localization approaches that leverage such tools therefore benefit from continuous improvements to SMT solvers.

Pavlinovic et al.\ developed MinErrLoc~\cite{Wies}, the state-of-the-art type error localization tool based on a variant of SMT called MaxSMT.
In the case of an ill-typed program, there is no satisfying assignment for the typechecking constraint problem.
from type inference. Instead, MinErrLoc seeks a minimum-weight set of constraints explaining
why no solution exists. Although effective at the time of its publication, MinErrLoc depends on a customized version of CVC4~\cite{CVC4}, rather than off-the-shelf MaxSMT solvers, and was not maintained after its original publication in 2014. % cite CVC4
Thus, MinErrLoc suffers from package rot and requires significant
effort to run. Our objectives were to bring the MinErrLoc approach up to modern standards,
and make it possible to leverage modern off-the-shelf MaxSMT solvers as originally intended.

Our main contribution is a new type error localization tool, Tyro,\footnote{\url{https://github.com/JKTKops/tyro}} inspired by the fundamental work of Pavlinovic et al. Tyro incorporates a new encoding for constraints resulting from polymorphic types, and is implemented with a two-stage design.
The first stage generates a human readable intermediate representation of the typechecking
constraint problem. Separate aspects of the problem are kept apart, increasing readability.
The second stage processes the intermediate representation into an SMT-LIB encoding~\cite{SMTLIB2-6},
bringing together separate aspects of the problem to form the encoded constraint system.
We also found that this architecture made the individual stages easier to debug, and therefore increases trust in the overall system.
Though the intermediate representation is specific to our system, we anticipate that the same ideas could be applied to a wide range of systems that leverage SMT.

Our experimental evaluation expands the evaluation of the MinErrLoc approach to a much larger dataset,
Validating the accuracy of Pavlinovic et al.'s approach, but also highlights the need for better
heuristics on some classes of programs.
We performed accuracy evaluations with a small expert-labeled dataset,
and both accuracy and performance evaluations with a large dataset automatically extracted from
student code in a large, introductory OCaml course.

\section{Overview of the MinErrLoc Approach}

Since our work builds on MinErrLoc~\cite{Wies}, a brief overview of its key ideas is warranted.\footnote{Overviews of OCaml's polymorphic types and of classical type inference for the system can be found in the Appendix.}
Primarily, we review the localization problem, the meaning of Pavlinovic et al.'s ``minimum error source'' heuristic,
and its reduction to MaxSMT.

Type errors result from (often minor) mistakes on the part of a programmer.
Correcting these mistakes will resolve the type error.
The program region containing the mistakes(s) is called the ``root cause'' of the type error.

The localization problem that we aim to solve is, given a program $P$ that
exhibits a type error, to identify the root cause.
This is an inherently ambiguous problem, because we cannot be certain exactly what the programmer intended.
The MinErrLoc approach follows Occam's Razor -- the simplest explanation is probably the correct one.

\subsection{Minimum Error Sources}\label{sec:mes}

An ``error source'' is a set of program locations which resolve the type error if removed from the program.\footnote{``Remove'' here means to replace by \lstinline!failwith "removed"!. Literally deleting the location would almost always result in syntax errors.} 
The root cause of the type error must be at a subset of an error source.

Not all error sources are equally likely to contain the true root cause, however. The MinErrLoc framework
provides an opportunity to specify a weight for every program location. A ``minimum error source'' is an
error source whose total weight is minimum. The framework allows these weights to be assigned independently
of constraint generation. Locations can also be set as ``hard constraints'' to tell the solver that they
should not be considered in potential error sources.

Consider this recursive OCaml program for finding the length of a list, which contains a bug:

\begin{lstlisting}
let rec len = function
  | [] -> 0.
  | _ :: xs -> 1 + len xs
\end{lstlisting}

This program is ill-typed, because the first arm produces a \lstinline!float!, but the second arm's use of \lstinline!+! means it produces an \lstinline!int!.
There is more than one way to explain this error. One possible error source is \lstinline!0.!, the float-valued first arm. Replacing this with an int-valued expression would resolve the error.
Another possible error source is the use of \lstinline!+!. Replacing \lstinline!+! with a \lstinline!float!-valued function could also resolve
the error.

If we use a trivial weighting heuristic, which simply assigns a weight of 1 to every location, then both error sources will be minimal.
However, domain knowledge might suggest that \lstinline!0.! is far more likely to be the true error source.
A weighting heuristic which considers the complexity of a program location, or which penalizes function calls,
might result in \lstinline!0.! as the unique minimum error source.

The MinErrLoc framework ensures that the constraint generation algorithm is independent of weight assignments.
This allows the framework to be re-used with different weighting heuristics.

\subsection{Reduction to MaxSMT}

MaxSMT is a variation of the SMT problem. Recall that SMT may be defined as the decision problem
asking whether a set $\mathcal C$ of propositional clauses is satisfiable. MaxSMT instead seeks a maximum
subset $\mathcal C' \subseteq \mathcal C$ such that $\mathcal C'$ is satisfiable. Note that maximizing the size
of $\mathcal C'$ corresponds to minimizing the size of $\mathcal C \setminus \mathcal C'$,
which will correspond to an error source. We may take this generalization of SMT two steps further.
First, we may include a \emph{weighting heuristic}, a function $w : \mathcal C \to \mathbb N$. Rather than
seeking a subset $\mathcal C'$ of maximum size, we seek a subset which maximizes $\sum_{c \in \mathcal C'} w(c)$.
This corresponds to the weighting heuristic for program locations mentioned above.
Finally, we may allow some clauses to be ``hard constraints,'' which \emph{must} be satisfied by the assignment.
The resulting problem is known as \emph{Partial Weighted MaxSMT}, but we will call it MaxSMT for brevity.

It would be easy to translate the typing constraints directly into MaxSMT constraints. Constraints in OCaml
programs are equality constraints between types. Equalities between (mono)types, and the types themselves,
can be encoded using the Theory of Inductive Datatypes~\cite{TheoryOfInductiveTypes},
which has been added to the SMT standard and is supported by SMT solvers such as Z3~\cite{SMTLIB2-6, Z3}.
A datatype (``sort'') is created in SMT which represents OCaml types. The OCaml types are then encoded
as values of this SMT datatype.

However, this encoding would not produce error sources - it would produce sets of typing constraints.
If several constraints arise from the same program point, the solver would be allowed to independently
decide whether or not to satisfy them. Instead, we must force the solver to decide on a location-by-location basis.
This is further complicated by the fact that the locations are not disjoint --
The location corresponding to an expression contains all of the locations corresponding to its subexpressions.
This tree structure is known as the \emph{abstract syntax tree}, or AST, of the program.

To accomplish this, weights are associated with program locations, rather than constraints.
The encoding of the constraints into MaxSMT incorporates information about the shape of the AST.
The encoding of the shape is such that removing a location also implicitly removes all of its children.
Otherwise we could be left with constraints to satisfy which are no longer in the program.
Our variation of the encoding is discussed in detail below.
%% Can't find a good place to include this, probably better to leave it off?
% This approach to encoding has a major downside for modularity and for tool development: the encoding is not
% a simple list of constraints. It is a flood of information about the AST shape and typing constraints together.
% This makes it difficult to read and debug, and difficult to re-use pieces of it in other tools which only care about the AST
% or only care about the constraints.

%% Putting the figure here places it at the top of the page about constraint generation.
\begin{figure*}[t]
  \begin{center}
  % Int
  \begin{bprooftree}
    \AxiomC{$\alpha$ new}
    \RightLabel{\textsc{Int}}
    \UnaryInfC{$\{\alpha =^\ell {\tt\bf int}\}; \Gamma \proves n^\ell : \alpha$}
  \end{bprooftree}
  % Bool
  \begin{bprooftree}
    \AxiomC{$\alpha$ new}
    \RightLabel{\textsc{Bool}}
    \UnaryInfC{$\{\alpha =^\ell {\tt\bf bool}\}; \Gamma \proves b^\ell : \alpha$}
  \end{bprooftree}
  % Var
  \begin{bprooftree}
    \AxiomC{$x : \forall \vec{\alpha}.(\mathcal C_x \Rightarrow \alpha_x) \in \Gamma$}
    \AxiomC{$\gamma,\vec{\beta}$ new}
    \RightLabel{\textsc{Var}}
    \BinaryInfC{$\{\gamma =^\ell \alpha_x, x^\ell(\vec{\beta})\}; \Gamma \proves x^\ell : \gamma$}
  \end{bprooftree}

  \bigskip

  % App
  \begin{bprooftree}
    \AxiomC{$\mathcal C_1; \Gamma \proves e_1 : \alpha$}
    \AxiomC{$\mathcal C_2; \Gamma \proves e_2 : \beta$}
    \AxiomC{$\gamma$ new}
    \RightLabel{\textsc{App}}
    \TrinaryInfC{$(\{\alpha =^\ell \fun{\beta}{\gamma}\} \cup \mathcal C_1 \cup \mathcal C_2);
      \Gamma \proves (e_1\ e_2)^\ell : \gamma$}
  \end{bprooftree}
  % Abs
  \begin{bprooftree}
    \AxiomC{$\mathcal C; \Gamma , x : \alpha_x \proves e : \beta$}
    \AxiomC{$\gamma$ new}
    \RightLabel{\textsc{Abs}}
    \BinaryInfC{$(\{\gamma =^\ell \fun{\alpha_x}{\beta}\} \cup \mathcal C);
      \Gamma \proves (\lambda x.e)^\ell : \gamma$}
  \end{bprooftree}

  \bigskip

  % Cond
  \begin{bprooftree}
    \AxiomC{$\mathcal C_1; \Gamma \proves e_1 : \alpha$}
    \AxiomC{$\mathcal C_2; \Gamma \proves e_2 : \beta$}
    \AxiomC{$\mathcal C_3; \Gamma \proves e_3 : \delta$}
    \AxiomC{$\gamma$ new}
    \RightLabel{\textsc{Cond}}
    \QuaternaryInfC{$
      (\{\alpha =^{\ell_1} {\tt\bf bool}, \beta =^{\ell_2} \gamma, \delta =^{\ell_3} \gamma\}
      \cup \mathcal C_1 \cup \mathcal C_2 \cup \mathcal C_3);
      \Gamma \proves
      \text{\lstinline!if! }
        e_1^{\ell_1}
      \text{ \lstinline!then! }
        e_2^{\ell_2}
      \text{ \lstinline!else! }
        e_3^{\ell_3}
      :
      \gamma 
    $}
  \end{bprooftree}

  \bigskip

  % Let
  \begin{bprooftree}
    \AxiomC{$\mathcal C_1; \Gamma \proves e_1 : \alpha_1$}
    \AxiomC{$
    \mathcal C_2; \Gamma , x : \forall\vec{\alpha}.(\mathcal C_1 \Rightarrow \alpha_1)
      \proves
    e_2 : \alpha_2
    $}
    \AxiomC{$\vec{\alpha} = fv(\alpha_1) \setminus fv(\Gamma)$}
    \AxiomC{$\vec{\beta},\gamma$ new}
    \RightLabel{\textsc{Let}}
    \QuaternaryInfC{$
        (\{\gamma =^\ell \alpha_2, x^\ell(\vec{\beta})\} \cup \mathcal C_2);
        \Gamma
      \proves (
        \text{\lstinline!let! }
          x = e_1
        \text{ \lstinline!in! }
          e_2
        )^\ell
      :
        \gamma
    $}
  \end{bprooftree}

  \end{center}

  \caption{Typing rules for the OCaml fragment}
  \label{fig:rules}
\end{figure*}

\section{Tyro Architecture}

%% Overview of the general architecture of Tyro PAGE 2.5 
%%  Frontend  
%%  Translation of typing constraints to intermediate language  (1 page)
%%   (fig. with sample program -> constraints incl. what they look like -> intermediate lang)
%%   Spec for intermediate language (simple grammar) 
%%  Translation of intermediate language to SMT  (1 column)
%%     (novel part : polymorphic types; modularity; ...)
%%  Backend  (1 column)
%%   MAXSat solver 

Tyro uses a modular, two-stage software architecture. The stages are implemented as
separate ``frontend'' and ``encoder'' tools. The input to the frontend is an OCaml program,
and the output is an Intermediate Representation of the constraints. The encoder accepts this IR,
and outputs an SMT-LIB script, which is then be passed to an off-the-shelf MaxSMT solver.

\subsection{Frontend}

The frontend's job is to extract a set of typing constraints from an OCaml program. We implemented
it by modifying EasyOCaml~\cite{EasyOCaml}. EasyOCaml is a tool with improved error message quality
for OCaml, and has also been modified for constraint generation in other work~\cite{Wies}.

First, a set of constraints are generated, including our representation of polymorphic types. Then, the
collected constraints are encoded into the intermediate representation.

The constraint generation is a modification of existing constraint-generation approaches~\cite{ConstraintForm,TypeInferenceConstrained,Wies}.
As a reminder, we focus on an idealized fragment of OCaml, shown in Figure~\ref{fig:fragment}.

\begin{figure}
\begin{flalign*}
  && \textbf{Expressions} && e ::&= x && \text{variable} && \\
  && && &\ \ |\ v && \text{value} && \\
  && && &\ \ |\ e\ e && \text{application} && \\
  && && &\ \ |\ \text{\lstinline!if! } e \text{ \lstinline!then! } e \text{ \lstinline!else! } e && \text{conditional} && \\
  && && &\ \ |\ \text{\lstinline!let! } x = e \text{ \lstinline!in! } e && \text{let binding} && \\
  \\
  && \textbf{Values} && v ::&= n && \text{integer} && \\
  && && &\ \ |\ b && \text{boolean} && \\
  && && &\ \ |\ \lambda x.e && \text{abstraction} && \\
  \\
  && \textbf{Monotypes} && \tau ::&= g\ |\ \alpha\ |\ \fun{\tau}{\tau} &&\ &&\\
  && \textbf{Polytypes} && \sigma ::&= \tau \ |\ \forall \alpha.\sigma &&\ &&
\end{flalign*}
  \caption{Idealized OCaml Fragment}
  \label{fig:fragment}
\end{figure}

The fragment supports variables, lambda abstraction, function application, conditionals, and local variable bindings.
The types $g$ are the ``ground types'', such as \texttt{int}, \texttt{float}, or \texttt{string}.
Types $\alpha$ represent globally unique type variables. These variables are \emph{monomorphic} -
they represent a single as-yet unknown type.
Polytypes, on the other hand, may universally quantify some or all of the variables in a monotype,
resulting in a template that can be re-used with multiple different types.

\subsection{Polymorphic Types}

Polymorphic types are a fundamental challenge for constraint-based type inference~\cite{ConstraintForm,EfficientGeneralization}.
When inferring a type for a polymorphic binding, a set of constraints will be generated. Some of these constraints will
refer to the polymorphic variables in the type of the binding. Whenever the binding is used, copies of these variables
are created in a process called \emph{instantiation}. Every copy of these variables must be independent from the others.
But every copy is also subject to the same constraints as the original. The solution taken by MinErrLoc is to also copy
all of the constraints. Our approach instead encodes these constraints as abstractions, allowing the MaxSMT solver decide when,
or indeed if, the copies should be created.

Since constraints associated with polytypes need to be recorded, a constraint set is attached to every polytype.
``Type schemes'' are a common approach to this in constraint-based systems~\cite{TypeInferenceConstrained,EssenceOfML,Wies}.
After inferring the type for a binding \lstinline!let! $x$ \lstinline!=! $e_1$ \lstinline!in! $e_2$, the variable $x$ will be added to the typing environment.
Its type will have the form: $$\forall \vec{\alpha}.(\mathcal{C}_x \Rightarrow \alpha_x)$$

\noindent where $\mathcal C$ is the associated set of constraints, and $\alpha_x$ is the type variable created for $e_1$.
We write simply $x : \alpha_x$ if $\vec{\alpha}$ and $\mathcal C_x$ are both empty.

When $x$ is later used, rather than create copies of the constraints in $\mathcal C_x$, we emit an ``instantiation constraint.''
These constraints are of the form $x(\bar{\beta})$, and have appeared previously in other Hindley-Milner-style systems~\cite{EssenceOfML}.
The constraint $x(\bar{\beta})$ represents the entire constraint set $\mathcal C_x[\vec{\beta}/\vec{\alpha}]$.
That is, the capture-avoiding substitution of the variables $\vec{\beta}$ for the variables $\vec{\alpha}$
in a copy of $C_x$.
Since instantiation constraints represent a set of regular typing constraints, they can appear
wherever a set of typing constraints can appear.

\subsection{Constraint Generation}

A typing constraint in Tyro takes the form $\tau_1 =^\ell \tau_2$. This is a simple equality between two types,
annotated with the program location $\ell$ where it was created. Since we need these locations to create the
constraints, we ensure that the AST nodes are annotated with locations as well.

Unlike MinErrLoc, our frontend does not encode the structure of the AST into the typing constraints. To improve
modularity and reusability, and to facilitate debugging, we keep this information separate for as long as possible.
This, along with instantiation constraints, simplifies the typing rules significantly. The rules are formulated
with a similar constraint typing relation, of the form: $$\mathcal C; \Gamma \proves e : \alpha$$

$\mathcal C$ is the set of constraints which have been emitted by inference for $e$.  $\Gamma$ is the typing environment
in which inference for $e$ should occur; $\Gamma$ maps variable names to type schemes. $e$ is a program expression,
and $\alpha$ is its inferred type.

Note that the relation always relates an expression to a type variable. This means that we cannot infer the type
\lstinline!int! for the expression \lstinline!0! - we must instead assign a new type variable $\alpha_0$ and emit
a constraint $\alpha_0 =$ \lstinline!int!. This prevents a loss of information. If we could infer the type \lstinline!int!
directly, and the expression \lstinline!0! were the root cause of the type error, there would be no link back to this
source location in the constraint set~\cite{Wies}. The typing rules are shown in Figure~\ref{fig:rules}.

Look in particular at the rules \textsc{Var} and \textsc{Let}, which are the main distinction from other
constraint-based systems. In the case of variables, we look up the type scheme from the environment.
Then we create new type variables to instantiate all variables in $\vec{\alpha}$.
However, we do not then copy $\mathcal C_x$. Instead, we emit an instantiation constraint
(with a location annotation). For let bindings, the difference is similar. Systems such as MinErrLoc
emit the entire constraint set $\mathcal C_1[\vec{\beta}/\vec{\alpha}]$ where we emit the instantiation
constraint $x(\vec{\beta})$. This instantiation constraint is necessary to ensure the consistency of
$\mathcal C_1$ - otherwise, if all uses of $x$ were removed from the program,
all constraints in $\mathcal C_1$ would be lost~\cite{ConstraintForm,Wies}.

%% end of section
The constraint generator is implemented as a modification of EasyOCaml~\cite{EasyOCaml}.
EasyOCaml is implemented as a fork of \lstinline!ocamlc!, the OCaml compiler. This unfortunately
pins it to a particular version of OCaml, which is not recent.
In order to support future work on newer versions of OCaml, we ported just the EasyOCaml constraint generation framework
to be a stand-alone OCaml project depending on the \lstinline!ocaml-base-compiler! package~\cite{BaseCompiler}.
Since this package does not include other features of EasyOCaml, it is significantly easier to port it to
new versions of OCaml.

\subsection{Intermediate Representation (IR)}

\begin{figure}[t]
  \small
  \begin{flalign*}
    && \textbf{Loc Index} && i ::=\ &n && && \\
    && \textbf{Weight} && \omega ::=\ &n && && \\
    && \textbf{Source Range} && \ell ::=\ &{line;col - line;col} && && \\
    && \textbf{Location} && L ::=\ &i\ \ell && \text{no weight given} && \\
    && && |\ &i\ \ell\ \omega && \text{weight given} && \\
    \\
    && \textbf{Constraint} && C ::=\ &i\ \tau_1 = \tau_2 && \text{equality} && \\
    && && |\ &i\ x(\vec{\beta}) && \text{instantiation} && \\
    \\
    && \textbf{Scheme} && S ::=\ &i\ x(\vec{\alpha})\ \vec{C} && \\
    && \textbf{IR} && R ::=\ &\vec{L}\ \vec{S}\ \vec{C} && \\
  \end{flalign*}

  \caption{IR Grammar}
  \label{fig:ir}
\end{figure}

The IR consists of three sets: a set of program source ranges, a set of type schemes,
and a set of constraints. Program locations may optionally be annotated by weights. Weights of zero correspond to
hard constraints. Whitespace is completely ignored. The complete expression grammar is shown in Figure~\ref{fig:ir}.
In constraints, $\tau$ refers to a monotype from Figure~\ref{fig:fragment}.

The ``Loc Indices'' $i$ must be distinct and essentially name the source ranges. Throughout the 
constraint (resp. schemes) portion of the IR, the indices are used to encode the source range
where the constraint (resp. schemes) was created. Later, the encoder will use the locations to embed
the shape of the AST into the encoding.

Each constraint scheme $S$ corresponds to a variable $x$ and its associated type scheme
$\forall \bar{\alpha}.(C_x \Rightarrow \tau_x$). In particular, the scheme relates the name $x$, the quantified variables $\bar{\alpha}$,
and the constraint set $C_x$. There is no special mention of $\alpha_x$.
The relationship between the scheme and $\alpha_x$ is encoded in how $\alpha_x$ (and its instantiations)
appear in the constraints. Regardless, for human readability, Tyro always places $\alpha_x$ at the end of $\bar{\alpha}$.

Every constraint is either an equality of OCaml monotypes (which can be type variables), or an instantiation constraint.
Instantiation constraints \textbf{can} appear inside schemes, which occurs whenever a polymorphic function is used
within a polymorphic definition.

Tyro generates the constraint portion of the IR from the constraint set $\mathcal C$ of
the top-level invocation of the constraint generation routine. Schemes are accumulated on the side, and always
emitted. Location annotations are treated similarly.

%%% This is a partially re-written section conclusion from the CAV paper I was working on.
%%% Something about it feels off.
The use of an intermediate representation is not necessary to the functionality of the system.
However, it offers several advantages. Primarily, unlike the SMT encoding, the IR
is human-writable and indeed human-readable given a bit of time. The final encoding, in contrast,
is deeply nested and littered with information about the AST structure, making it quite difficult to read or write.
Inspecting these intermediate files was invaluable for debugging constraint generation, and writing
them by hand was further valuable for debugging the SMT encoder. This separation makes it easier to trust
the correctness of the constraint generation and encoding steps.

Additionally, the use of an IR promoted modularity and reusability between
the components. While working on Tyro, we were able to mix-and-match different methods of encoding the IR,
without making any changes at all to the constraint generator.
Similarly, we were able to redesign a significant portion of
the constraint generator without any fear of breaking the encoder.

\subsection{SMT Encoder}

\begin{figure*}
  \newcommand{\located}[1]{^{\ell_{#1}}}

  \begin{subfigure}[t]{0.50\textwidth}
    \begin{lstlisting}[mathescape,numbers=none]
      let$\located{0}$ x = "hi"$\located{1}$ in (not$\located{2}$ x$\located{3}$)$\located{4}$
    \end{lstlisting}
    \caption{}
  \end{subfigure}
  \quad
  \begin{subfigure}[t]{0.47\textwidth}
    \Tree[.{(\lstinline!let x = "hi"$\ $in not x!)$^{\ell_0}$}
        {\lstinline!"hi"!$^{\ell_1}$}
        [.{(\lstinline!not x!)$^{\ell_4}$}
          {\lstinline!not!$^{\ell_2}$}
          {\lstinline!x!$^{\ell_3}$} ]]

    \caption{}
    \label{fig:ex:locs}
  \end{subfigure}

  \begin{subfigure}{0.47\textwidth}
  \centering
  \begin{varwidth}{\linewidth}
\begin{verbatim}
<locations omitted>
---
0 x('x) {
  1 'x = string
}
---
0 x('x0)
2 'l2 = bool -> bool
3 x('x1)
4 'l2 = 'x1 -> 'l4
\end{verbatim}
  \end{varwidth}
  \caption{}
  \label{fig:ex:transl:ir}
  \end{subfigure}
  \begin{subfigure}{0.6\textwidth}
    \centering
    \small
    \begin{varwidth}{\linewidth}
\begin{lstlisting}[mathescape,language=lisp,keywordstyle=\ttfamily,numbers=none]
(declare-datatype Type
  ((string) (bool) (-> (->.1 Type) (->.2 Type))))
(declare-const $\ell_0$ Bool)(assert-soft $\ell_0$ :weight 5)
(declare-const $\ell_1$ Bool)(assert-soft $\ell_1$ :weight 1)
...
(declare-const -x0 Type)(declare-const -l2 Type) ...
(define-fun x ((-x Type)) Bool
  (=> $\ell_0$ (=> $\ell_1$ (= -x string))))
(assert
  (=> $\ell_0$ (and (x -x0)
    (=> $\ell_4$ (and (= -l2 (-> -x1 -l4))
      (=> $\ell_2$ (= -l2 (-> bool bool)))
      (=> $\ell_3$ (x -x1)))))))
(check-sat)(get-objectives)(get-value ($\ell_0\ \ell_1\ \ell_2\ \ell_3\ \ell_4$))
\end{lstlisting}
  \end{varwidth}
  \caption{}
  \label{fig:ex:transl:smt}
  \end{subfigure}

  \caption{
    A sample run of Tyro.\newline
      (a) an ill-typed program from~\cite{Wies} with locations annotated;
      (b) labeled program AST;\newline
      (c) simplified intermediate representation;
      (d) SMT encoding.
  }
  \label{fig:ex:transl}
\end{figure*}

The SMT encoding step translates the intermediate representation to SMT-LIB~\cite{SMTLIB2-6} code.
The only extension required to SMT-LIB 2.6 is vZ, for MaxSMT~\cite{vZ}.
A Tyro run on the example from Section 2.2 of \cite{Wies}
can be seen in Figure~\ref{fig:ex:transl}. In particular, our SMT encoding
is in Figure~\ref{fig:ex:transl:smt}.

Type schemes become SMT interpreted functions for the solver to instantiate on-demand. Equality constraints
on types are encoded directly as equality constraints in the theory of inductive datatypes, using a
\texttt{Type} sort to represent OCaml types. The \texttt{Type} sort is as described for MinErrLoc~\cite{Wies}.

Type variables are encoded with a ``-'' in front of their name, to avoid conflicts with scheme names.
This serves the same purpose as the single quote (``tick'') in OCaml source code,
but ticks are not allowed at the start of an SMT variable name.

The SMT encoding of constraints incorporates information about the AST structure.
The enumeration of source locations is examined to recover an ``AST forest.'' Each interval
in the enumeration becomes a (possibly indirect) child of every interval that contains it.
The result is a forest of program locations. In practice, this forest contains one tree for
every top-level expression or let binding or in the program.

Consider the program fragment:
\begin{displaymath}
  \text{\lstinline!let x = "hi"$\ $in not x!}
  \tag{Ex.}\label{ex}
\end{displaymath}

There are 5 source ranges in this fragment, shown in Figure~\ref{fig:ex:locs}.
If the MaxSMT solver decides to remove the entire fragment (location $\ell_0$, the root of the tree),
then all four of the other subfragments are necessarily removed as well.
The weight of this decision must be determined only by the weight of location $\ell_0$,
even though all of its children are also being removed.

Therefore, for the fragment above, we encode a constraint $C$ at location $\ell_3$ as
$$\ell_0 \Rightarrow (\ell_4 \Rightarrow (\ell_3 \Rightarrow C))$$

The location variables $\ell_i$ are (softly) asserted directly with their weight. For example, with this
fragment, we have

\begin{center}
\begin{tabular}{c} % needed to center a listing, apparently
\begin{lstlisting}[mathescape,language=lisp,keywordstyle=\ttfamily,numbers=none]
(assert-soft $\ell_0$ :weight 5)
(assert-soft $\ell_3$ :weight 1)
(assert-soft $\ell_4$ :weight 3)
\end{lstlisting}
\end{tabular}
\end{center}

The decision to remove location $\ell_0$ (by setting the SMT variable $\ell_0$ to \lstinline!false!)
now carries a cost of 5. The constraint $C$ would no longer be active, even if $\ell_3$ and $\ell_4$
were still set to \lstinline!true!.

All constraints are encoded in this way, starting at the root of an AST.
Paths are combined, such that all of the constraints associated with a particular top-level
statement are encoded into a single \texttt{assert} form.
For example, two constraints $C_1$, $C_2$ at location $\ell_4$ would be
represented by only one copy of the above constraint encoding, with $C$ = $C_1 \wedge C_2$. We apply this
in a nested fashion, so each assertion consists of many nested implications and constraints.
The constraints contained in a type scheme are also encoded this way, but are placed into an
SMT ``defined function'' rather than using an \texttt{assert} form.
The assertion tree for a scheme is rooted at the AST node which defined the scheme.
In the case of a distant reference to a let-bound variable, this ensures that the instantiation constraint's
implied constraints are disabled if (any parent of) the let binding is removed.\footnote{
Instantiation constraints for a scheme can arise in only two cases:
the binding is local, and the reference is a child of the binding in the AST; or
the binding is top-level, and therefore the scheme's root is also the AST root.}

The encoder provides MinErrLoc's weighting heuristic as a default if weights are not provided.
Each node in the AST forest is assigned a weight equal to the size of the sub-AST rooted at that node.
In example~(\ref{ex}) above, location 4 is assigned weight 3, ensuring that removing location 4 is more costly
than removing both location 2 and 3 (which have a cumulative weight of 2).
In its current configuration, Tyro uses the default weight for almost all locations.
% mention exceptions? There are two:
%   references to library functions are effectively set as hard.
%     What actually happens is that type judgement will look like
%     the type of the library function is a known constant, same as if
%     we had typechecked an 'int' or 'bool' etc.
%     This has the same effect as creating a synthetic location and equating the
%     library type with the type variable using the synthetic location, then setting
%     that location to be hard. In fact, before I changed it, EasyOCaml was
%     generating such locations.
%  Locations corresponding to partial applications are always set to be hard.
%    According to comments in the MinErrLoc artifact (presumably due to Pavlinovic),
%    situations where these partial applications were reported were "not useful" and
%    a message corresponding to either the entire application or just the application
%    head made more sense.

Inspecting Figure~\ref{fig:ex:transl:ir}, the IR illustrates the
change from MinErrLoc's encoding to Tyro's:
the constraint \lstinline!'x = string! came from a let binding, and is now part of a scheme.
When the script in Figure~\ref{fig:ex:transl:smt} is run through Z3~\cite{Z3},
location $\ell_1$ is identified as the error source.\footnote{There are 3 minimal error sources
for this program: $\{\ell_1\}, \{\ell_2\},$ and $\{\ell_3\}$.}
%If a user does not believe $\ell_1$ to be the error, they can modify it to be a hard constraint
%and then retry.}% footnote worth having? -- tried cutting it shorter instead

%% be more specific and/or move this to evaluation section?
Taking advantage of the modularity offered by our design, we also implemented another SMT encoding
which avoids deeply-nested implications. The shallower encoding appears to help the SMT solver in some cases.
When the minimal cost is high, the shallower encoding can result in error sources that are not actually minimal.
Empirically, however, almost all error sources for programs in our dataset had low costs.
The MinErrLoc artifact employs the same alternate encoding, so we used it while evaluating Tyro.

\subsection{Backend}
% Overview predicted this would take a column, but I don't really have anything else
% to say about it. Maybe there's something I'm not thinking of?

The output of the encoder is an SMT-LIB script. The scripts are compatible with any SMT solver
that supports at least SMT-LIB 2.6~\cite{SMTLIB2-6} and the vZ extension for MaxSMT~\cite{vZ}.
Tyro uses Z3 by default. The output of the SMT solver is processed to extract the
minimum error source.

\section{Evaluation}

The MinErrLoc approach was evaluated for performance on a dataset
of 356 programs collected from a programming course~\cite{Wies}.
We collected several thousand programs from a programming course~\cite{DataCollection} and took a random sample
of 500 programs each from three different assignments for a total of 1500 programs.
Programs were only selected if they could be parsed, but did not compile.
Of the 1500 programs selected, approximately 70 contained localized errors other than type mismatches and were discarded.
As Tyro is an experiment in delayed instantiation, we focused our evaluation on delayed instantiation.
Though constraint slicing and preemptive cutting are shown to be both effective and simple to
implement by MinErrLoc, our evaluation of Tyro did not use them.

\subsection{Timing}

Our statistics for timing Tyro are shown in Figure~\ref{fig:timing}. Experiments were conducted on
an Intel(R) Core(TM) i7-8550U CPU with four 1.80 GHz cores. Our experiments only used a single
core for each instance of Tyro, but ran Tyro on several programs simultaneously. Tyro was run with
a 100 second timeout, which excluded a further 40 programs, all from the same homework assignment.
The statistics shown are for the remaining 1388 programs, in a format easily compared with
MinErrLoc's evaluation in Figure 11 of Pavlinovic et al.~\cite{Wies}.

We split our dataset into groups based on program length in lines of code. The number in parentheses is
the number of programs in that group.
The number of equality constraints, the minimum error source weight, and the time
to run Tyro were recorded for each program. Note that the number of equality constraints cannot indicate
how many times instantiation constraints will cause those equality constraints to be copied; therefore it
is only a lower bound on the complexity of the MaxSMT problem.

In all groups, the constraint counts generated for our programs are significantly higher than those for
MinErrLoc's evaluation. This suggests a difference in the typical structure of the programs which makes
the evaluations hard to compare. Despite the slower processor used in our experiments and the generally
higher constraint counts, we exhibit remarkably similar minimum and median execution times. Approximately
2.9\% of programs evaluated timed out, and our maximum execution times are similar, though again slower,
to those of MinErrLoc's evaluation for groups with similar constraint counts.

Our results are therefore promising. Our evaluation largely affirms that of MinErrLoc, on a significantly larger dataset.

One potential explanation for the lack of significant improvement is to consider how the SMT solver
proceeds with instantiation constraints. As noted by MinErrLoc, the time spent copying constraint sets
for instantiation during constraint generation is significant~\cite{Wies}. By delaying this work to the
SMT solver, we create opportunities for the solver to recognize that an instantiation is not necessary at all.
But we also risk that the SMT solver may perform a single instantiation many times.
Given the cost of instantiations, the risks may outweigh the benefits for the version of Z3 used.
This may improve in the future as solvers improve. We posit that SMT scripts generated by Tyro may make
good benchmarks for MaxSMT solvers.

\begin{figure*}
  \centering
  \begin{tabular}{ |c r|c|c|c|c|c|c|c|c|c| }
    \hline
    \multicolumn{2}{|c|}{\multirow{2}{*}{Group}}
    & \multicolumn{3}{|c|}{Constraints} & \multicolumn{3}{|c|}{Weight} & \multicolumn{3}{|c|}{Time (s)} \\
    \cline{3-11}
    & & \textbf{min} & \textbf{med} & \textbf{max} & \textbf{min} & \textbf{med} & \textbf{max} & \textbf{min} & \textbf{med} & \textbf{max} \\
    \hline
    0-50 & (5) & 44 & 63 & 72 & 1 & 1 & 3 & 0.02 & 0.11 & 0.16 \\
    \hline
    50-100 & (57) & 96 & 276 & 990 & 1 & 2 & 35 & 0.08 & 0.68 & 2.93 \\
    \hline
    100-150 & (659) & 111 & 532 & 1741 & 1 & 2 & 33 & 0.09 & 2.62 & 85.18 \\
    \hline
    150-200 & (449) & 399 & 976 & 2341 & 1 & 3 & 23 & 0.84 & 17.80 & 87.86 \\
    \hline
    200-250 & (55) & 696 & 1463 & 2702 & 1 & 2 & 18 & 1.53 & 10.50 & 89.43 \\
    \hline
    250-300 & (13) & 633 & 1514 & 3039 & 1 & 1 & 6 & 2.94 & 7.58 & 86.31 \\
    \hline
    300-350 & (5) & 1073 & 1516 & 2690 & 1 & 2 & 3 & 8.52 & 14.10 & 50.44\\
    \hline
  \end{tabular}
  
  \caption{Statistics for Tyro execution on whole programs}
  \label{fig:timing}
\end{figure*}

\begin{figure}
  \centering
  \begin{tabular}{ |c|c||c| }
    \hline
    Tyro & OCaml & \# of outcomes \\
    \hline
    hit & hit & 5 \\
    \hline
    hit & close & 6 \\
    \hline
    hit & miss & 3 \\
    \hline
    close & hit & 3 \\
    \hline
    close & close & 20 \\
    \hline
    close & miss & 1 \\
    \hline
    miss & hit & 3 \\
    \hline
    miss & close & 0 \\
    \hline
    miss & miss & 1 \\
    \hline
  \end{tabular}
  \caption{Accuracy on expert-labeled programs}
  \label{fig:acc:hand}
\end{figure}

\begin{figure}
  \centering
  \begin{tabular}{ |c|c||c| }
    \hline
    Tyro & OCaml & \# of outcomes \\
    \hline
    hit & hit & 430 \\
    \hline
    hit & close & 9 \\
    \hline
    hit & miss & 15 \\
    \hline
    close & hit & 39 \\
    \hline
    close & close & 11 \\
    \hline
    close & miss & 2 \\
    \hline
    miss & hit & 113 \\
    \hline
    miss & close & 3 \\
    \hline
    miss & miss & 25 \\
    \hline
  \end{tabular}
  \caption{Accuracy on automatically labeled programs}
  \label{fig:acc:auto}
\end{figure}

\newcommand{\mycomment}[1]{}
\mycomment{
0-50 (5): cstrs:[44, 63.0, 72] time:[0.02, 0.11, 0.16] weight:[1, 1.0, 3]
50-100 (57): cstrs:[96, 276.0, 990] time:[0.08, 0.68, 2.93] weight:[1, 2.0, 35]
100-150 (659): cstrs:[111, 532.0, 1741] time:[0.09, 2.62, 85.18] weight:[1, 2.0, 33]
150-200 (449): cstrs:[399, 976.0, 2341] time:[0.84, 17.8, 87.86] weight:[1, 3.0, 23]
200-250 (55): cstrs:[696, 1463.0, 2702] weight:[1, 2.0, 18] time:[1.53, 10.5, 89.43] 
250-300 (13): cstrs:[633, 1514.0, 3039] weight:[1, 1.0, 6] time:[2.94, 7.58, 86.31] 
300-350 (5): cstrs:[1073, 1516.0, 2690] weight:[1, 2.0, 3] time:[8.52, 14.1, 50.44]
}
\mycomment{
Summary {row1 = (Hit,Hit,430), row2 = (Hit,NearMiss,9), row3 = (Hit,Miss,15), row4 = (NearMiss,Hit,39), row5 = (NearMiss,NearMiss,11), row6 = (NearMiss,Miss,2), row7 = (Miss,Hit,113), row8 = (Miss,NearMiss,3), row9 = (Miss,Miss,25)}
Summary {row1 = (Hit,Hit,5), row2 = (Hit,NearMiss,6), row3 = (Hit,Miss,3), row4 = (NearMiss,Hit,3), row5 = (NearMiss,NearMiss,20), row6 = (NearMiss,Miss,1), row7 = (Miss,Hit,3), row8 = (Miss,NearMiss,0), row9 = (Miss,Miss,1)}
}

\subsection{Localization Accuracy}\label{sec:tyro:localization-acc}

We first took a random sample of 50 programs from our data set and labeled the
true error source by hand. 8 of the programs were discarded because we could not
decide which of several error sources were most likely.
The comparison of Tyro's accuracy versus \lstinline!ocamlc!'s on these programs is shown in
Figure~\ref{fig:acc:hand}. They are formatted for easy comparison to Figure 8 of
the MinErrLoc analysis~\cite{Wies}.
Regions were marked as ``hit'' if they exactly matched the true error source.
If the region was close enough for a (novice) programmer to easily understand the
true problem, the region was marked as ``close.'' Otherwise, it is marked ``miss.''

Our expert-labeled evaluation uses a larger dataset than MinErrLoc's expert-labeled
evaluation (40 programs versus 20) and displays almost identical proportions of outcomes.
This reaffirms the small-scale evaluation results of MinErrLoc.

We reviewed the one program where both Tyro and OCaml missed.
It is an especially tricky case where the true error source contains two program locations,
and their relationship is partially obscured by the programmer's mistake. Tyro and OCaml
report adjacent program locations (both of weight 1), neither of which are members of the true
error source.\footnote{
However, if both OCaml and Tyro's reported locations are made hard constraints,
the true error source becomes a minimum error source.
}
In the other 41 programs, either Tyro or OCaml identify the true error source.

We experimented with automatic methods for evaluating localization accuracy,
using a similar approach to~\cite{DataDriven}.
We compare the region(s) reported by localization to the region(s)
that students actually modified to fix a type error.
For each of the 1388 programs in our random sample, we determined if the successive code sample
from the same student compiled successfully.
We recover the regions that the student modified using Difftastic~\cite{difftastic},
a structural differencing tool, and then removed programs where Difftastic reported
a high portion of the file had been rewritten. In this manner, we collected 647 data points.
We then classified the identified regions in an
automated manner similar to the expert-labeled evaluation. Exact matches were marked
as ``hit'', other forms of (possibly partial) overlap or shared endpoints were
marked as ``close'', and anything else was marked as a ``miss.''
Notably, consider an application such as \lstinline!f x!.
If the student modified \lstinline!x!, but the identified region was \lstinline!f!,
these intervals are considered to share an endpoint and are marked ``close.''
This situation appears to be quite common, as does the reverse.

Unfortunately, this approach suffers from a major source of bias: because the students
fixing the program only had access to error messages from OCaml, they were far more likely
to modify the region of code indicated by OCaml (which is always a member of some error source).
This bias is clearly seen in the results in Figure~\ref{fig:acc:auto}.

As part of typical homework assignments in our course, students write their own test cases.
These test cases are formatted as lists of input-output pairs. One test case was part of the given code.
For some problems, the given test case was correct.
For other problems, students were supposed to fix an incorrect test case.
We inspected a random sample of the 113 programs where Tyro missed but OCaml hit.
In approximately 70\% of the sampled programs, the type error was due to malformed test cases.
The students wrote several test cases containing \lstinline!int!s where \lstinline!float!s were expected, or vice versa.
Because the students wrote several cases after the one given case, the minimum error source
is always the given test case.
But the given test case comes first, so OCaml reports the mismatch on the cases written by the student.
Tyro ``misses'' for these programs because the students followed OCaml's advice --
even when that advice was incorrect.

This demonstrates the subjectivity of the type error localization problem,
and provides evidence that type annotations should be used judiciously to guide students.
If a top-level type annotation had been included for the test cases and set as a hard location,
Tyro and OCaml would both identify the incorrect test cases.\footnote{Such annotations are recommended by Pavlinovic et al.~\cite{Wies}, but
unfortunately we did not have control over the content of the assignments.}

Considering this bias, Tyro appears remarkably accurate despite the fact that we are using the
``relatively simplistic'' weighting heuristic of AST size. This again reaffirms the potential
of the MaxSMT localization approach.

Out of the 647 programs evaluated, either Tyro or OCaml identify the true error source in over 96\% of cases.
This is similar to our observation from the expert-labeled evaluation. Therefore, we conclude that reporting
localizations from Tyro alongside OCaml's error report would be an effective, accurate diagnostic for programmers.

\section{Related Work}
%% A draft related work section that I wrote when I started working on the paper for CAV.
%% Parts of it might still be useful here?

MinErrLoc~\cite{Wies} first demonstrated that type error localization problems can be
efficiently expressed as Partial Weighted MaxSMT problems.
They recognize the issues associated with polymorphic types, but do not simplify them.
They propose two algorithms to improve the situation: Lazy Quantifier-Based Instantiation,
and Lazy Unification-Based Instantiation. Tyro implements Lazy Quantifier-Based Instantiation.

% Typpete really feels like a typo of Typette, but it is the name given in their paper.
Other tools have also begun using (Max)SMT solvers for type inference problems.
Typpete~\cite{Typpete} uses a MaxSMT solver to infer type annotations to be added to Python programs.
Typpete additionally had to solve the challenge of encoding subtyping constraints.
Similar ideas were discussed in the presentation of MinErrLoc.
We believe our architecture could be leveraged to tie these ideas together
and create localization tools for languages like Java or Haskell.

Mycroft~\cite{Mycroft} takes a different approach to localization by heuristic minimization.
Rather than reducing localization to MaxSMT, Mycroft is a solver dedicated to minimizing
error sources in type inference problems. It is generalized over the type system being used and
requires an inference engine for that system. The Mycroft algorithm is very similar to
MaxSMT algorithms based on ``Unsatisfiable Cores''~\cite{UnsatCores}. Mycroft's ability to use
a dedicated typechecking engine means it can avoid issues like the polymorphic constraint blowup
seen in MinErrLoc and Tyro. Unfortunately, Mycroft does not benefit from frequent improvements
to the MaxSMT state-of-the art.

Zhang and Myers have previously reduced localization problems to finding
certain types of paths in a graph~\cite{TowardsGeneralDiagnosis}. They apply Bayesian methods
to guess which source location to blame for the faulty paths. 
This work was further developed to support advanced type system features like \emph{type classes}
in Haskell~\cite{DiagnosingWithClass} and an implementation, SHErrLoc, is available~\cite{SHErrLoc}.
Their graphs did not encode the ``flow'' of typing information during the inference process.
A recent approach, HM$^\ell$, takes inspiration from subtyping systems to express the way
that typing information flows through the inference process~\cite{GettingIntoTheFlow}.
Rather than heuristically producing a localization guess, HM$^\ell$ error messages contain
a detailed flow diagram containing all of the source locations participating in the error.
They report that this can lead to ``information overload,'' however, it is a promising new
view on the problem.

\section{Future Work}

We have observed several potential avenues for future work on Tyro or other tools.
The most obvious is perhaps to improve the weighting heuristic.

While Tyro implements the Lazy Quantifier-Based Instantiation proposal from~\cite{Wies},
a unification-based algorithm was also proposed.
The proposed algorithm makes several calls to the SMT solver,
and requires changing the constraints related to polymorphic variables on every call to the solver.
This would be a considerable challenge for the architecture of MinErrLoc.
However, because Tyro separates constraints related to polymorphic variables from other constraints,
it seems the algorithm could be implemented on top of Tyro in a relatively straightforward fashion,
which we intend to explore in future work.

For future work on MaxSMT solvers, we believe that MaxSMT scripts generated by
Tyro have potential as benchmarks.

\section{Conclusion}

% My guess from what I've seen in the data is that this improvement is not enough to
% catch up to Mycroft, which is not surprising, especially on larger problems.
% Mycroft loses out on the constant improvements to SMT solvers, but more than makes
% up for the difference by allowing a type solver dedicated to the type system
% in question.
% Potential future work proposal: help the SMT solver significantly by using
% a custom theory solver dedicated to the "theory of OCaml types."
% This would be an SMT analogue of the Mycroft work.

Tyro is a modernization of the MinErrLoc MaxSMT approach to type error localization.
Our evaluation reaffirms the accuracy and performance potential of the approach using
a larger dataset. Our evaluation for accuracy indicates that a less simplistic metric
than AST size may perform better, at least on student programs.
Regardless, our evaluation shows that the combination of Tyro and OCaml already exhibits
an accuracy above 96\%.

Tyro's modular design makes it easy to experiment with modifications to various aspects
of the system. Indeed, we experimented with some variations on the AST size heuristic,\footnote{MinErrLoc also incorporates at least one such variation.}
completely rewriting the constraint generation frontend, and several SMT encodings.
While incorporating lazy quantifier-based instantiation did not immediately improve the
performance of the approach, we believe Tyro's architecture will allow it to serve as a
testbed for future work on MaxSMT-based localization.

\section{Acknowledgements}

This work was supported by the Social Sciences and Humanities Research Council (SSHRC), the OCaml Software Foundation,
and the Canada CIFAR AI Chair Program.

\bibliographystyle{IEEEtran}
\bibliography{cleaned}

\begin{thebibliography}{10}
\providecommand{\url}[1]{#1}
\csname url@samestyle\endcsname
\providecommand{\newblock}{\relax}
\providecommand{\bibinfo}[2]{#2}
\providecommand{\BIBentrySTDinterwordspacing}{\spaceskip=0pt\relax}
\providecommand{\BIBentryALTinterwordstretchfactor}{4}
\providecommand{\BIBentryALTinterwordspacing}{\spaceskip=\fontdimen2\font plus
\BIBentryALTinterwordstretchfactor\fontdimen3\font minus \fontdimen4\font\relax}
\providecommand{\BIBforeignlanguage}[2]{{%
\expandafter\ifx\csname l@#1\endcsname\relax
\typeout{** WARNING: IEEEtran.bst: No hyphenation pattern has been}%
\typeout{** loaded for the language `#1'. Using the pattern for}%
\typeout{** the default language instead.}%
\else
\language=\csname l@#1\endcsname
\fi
#2}}
\providecommand{\BIBdecl}{\relax}
\BIBdecl

\bibitem{OCaml}
\BIBentryALTinterwordspacing
{OCaml Foundation}, ``{OCaml}.'' [Online]. Available: \url{https://ocaml.org/}
\BIBentrySTDinterwordspacing

\bibitem{HowStudentsFixedTypeErrors}
B.~Wu and S.~Chen, ``How type errors were fixed and what students did?'' in \emph{Proceedings of the ACM on Programming Languages}, vol. 1, OOPSLA, Oct. 2017, pp. 105:1--27.

\bibitem{SearchingForErrors}
B.~S. Lerner, M.~Flower, D.~Grossman, and C.~Chambers, ``Searching for type-error messages,'' in \emph{Proceedings of the 28th ACM SIGPLAN Conference on Programming Language Design and Implementation}, Jun. 2007, p. 425–434.

\bibitem{Helium}
B.~Heeren, D.~Leijen, and A.~van IJzendoorn, ``Helium, for learning {Haskell},'' in \emph{Proceedings of the 2003 {ACM} {SIGPLAN} workshop on {Haskell}}, Aug. 2003, pp. 62--71.

\bibitem{SPS}
M.~Wand, ``A semantic prototyping system,'' in \emph{Proceedings of the 1984 SIGPLAN Symposium on Compiler Construction}, Jun. 1984, p. 213–221.

\bibitem{FindingTheSource}
M.~Wand, ``Finding the source of type errors,'' in \emph{Proceedings of the 13th {ACM} {SIGACT}-{SIGPLAN} {Symposium} on {Principles} of {Programming} {Languages}}, Jan. 1986, pp. 38--43.

\bibitem{GettingIntoTheFlow}
I.~Bhanuka, L.~Parreaux, D.~Binder, and J.~I. Brachthäuser, ``\BIBforeignlanguage{en}{Getting into the {Flow}: {Towards} {Better} {Type} {Error} {Messages} for {Constraint}-{Based} {Type} {Inference}},'' in \emph{\BIBforeignlanguage{en}{Proceedings of the ACM on Programming Languages}}, vol. 7, OOPSLA2, Oct. 2023, pp. 431--459.

\bibitem{LearningToBlame}
E.~L. Seidel, H.~Sibghat, K.~Chaudhuri, W.~Weimer, and R.~Jhala, ``Learning to {Blame}: {Localizing} {Novice} {Type} {Errors} with {Data}-{Driven} {Diagnosis},'' in \emph{Proceedings of the ACM on Programming Languages}, vol. 1, OOPSLA, Oct. 2017.

\bibitem{NoviceTypeErrorDiagnosis}
C.~Geng, H.~Ye, Y.~Li, T.~Han, B.~Pientka, and X.~Si, ``\BIBforeignlanguage{en}{Novice {Type} {Error} {Diagnosis} with {Natural} {Language} {Models}},'' in \emph{\BIBforeignlanguage{en}{Programming {Languages} and {Systems}}}, Dec. 2022, pp. 196--214.

\bibitem{ConstraintForm}
M.~Sulzmann, M.~Muller, and C.~Zenger, ``Hindley/{Milner} style type systems in constraint form,'' Tech. Rep., Oct. 1999.

\bibitem{EssenceOfML}
F.~Pottier and D.~Rémy, ``{The Essence of ML Type Inference},'' in \emph{Advanced Topics in Types and Programming Languages}, Jan. 2005, pp. 389--489.

\bibitem{EfficientGeneralization}
\BIBentryALTinterwordspacing
O.~Kiselyov, ``Efficient and {Insightful} {Generalization}.'' [Online]. Available: \url{https://okmij.org/ftp/ML/generalization.html}
\BIBentrySTDinterwordspacing

\bibitem{TypeInferenceConstrained}
M.~Odersky, M.~Sulzmann, and M.~Wehr, ``\BIBforeignlanguage{en}{Type inference with constrained types},'' \emph{\BIBforeignlanguage{en}{Theory and Practice of Object Systems}}, vol.~5, no.~1, pp. 35--55, Jan. 1999.

\bibitem{Mycroft}
C.~Loncaric, S.~Chandra, C.~Schlesinger, and M.~Sridharan, ``{A Practical Framework for Type Inference Error Explanation},'' in \emph{Proceedings of the 2016 ACM SIGPLAN International Conference on Object-Oriented Programming, Systems, Languages, and Applications}, vol.~51, no.~10, Oct. 2016, p. 781–799.

\bibitem{Z3}
L.~De~Moura and N.~Bj\o{}rner, ``{Z3: An Efficient SMT Solver},'' in \emph{Proceedings of the Theory and Practice of Software, 14th International Conference on Tools and Algorithms for the Construction and Analysis of Systems}, Mar. 2008, p. 337–340.

\bibitem{Wies}
Z.~Pavlinovic, T.~King, and T.~Wies, ``Finding minimum type error sources,'' in \emph{Proceedings of the 2014 {ACM} {International} {Conference} on {Object} {Oriented} {Programming} {Systems} {Languages} \& {Applications}}, Oct. 2014, pp. 525--542.

\bibitem{CVC4}
C.~W. Barrett, C.~L. Conway, M.~Deters, L.~Hadarean, D.~Jovanovic, T.~King, A.~Reynolds, and C.~Tinelli, ``{CVC4},'' in \emph{Proceedings of Computer Aided Verification - 23rd International Conference}, vol. 6806, Jul. 2011, pp. 171--177.

\bibitem{SMTLIB2-6}
\BIBentryALTinterwordspacing
C.~Barrett, P.~Fontaine, and C.~Tinelli, ``{The SMT-LIB Standard: Version 2.6},'' Department of Computer Science, The University of Iowa, Tech. Rep., 2017. [Online]. Available: \url{www.SMT-LIB.org/papers/smt-lib-reference-v2.6-r2021-05-12.pdf}
\BIBentrySTDinterwordspacing

\bibitem{TheoryOfInductiveTypes}
C.~Barrett, I.~Shikanian, and C.~Tinelli, ``{An Abstract Decision Procedure for a Theory of Inductive Data Types},'' \emph{Journal on Satisfiability, Boolean Modeling, and Computation (JSAT)}, vol.~3, pp. 21--46, Jul. 2007.

\bibitem{EasyOCaml}
\BIBentryALTinterwordspacing
B.~Becker, C.~Haack, and J.~B. Wells, ``{EasyOCaml}.'' [Online]. Available: \url{http://easyocaml.forge.ocamlcore.org/}
\BIBentrySTDinterwordspacing

\bibitem{BaseCompiler}
\BIBentryALTinterwordspacing
X.~Leroy, ``ocaml-base-compiler.'' [Online]. Available: \url{https://ocaml.org/p/ocaml-base-compiler/}
\BIBentrySTDinterwordspacing

\bibitem{vZ}
N.~Bj{\o}rner and P.~Dung, ``{vZ - Maximal Satisfaction with Z3},'' in \emph{Proceedings of the 6th International Symposium on Symbolic Computation in Software Science}, Dec. 2014.

\bibitem{DataCollection}
A.~Ceci, H.~C.~A. Tavante, B.~Pientka, and X.~Si, ``Data {Collection} for the {Learn-OCaml} {Programming} {Platform}: {Modelling} {How} {Students} {Develop} {Typed} {Functional} {Programs},'' in \emph{{SIGCSE} '21: The 52nd {ACM} Technical Symposium on Computer Science Education}, Mar. 2021, p. 1341.

\bibitem{DataDriven}
\BIBentryALTinterwordspacing
E.~L. Seidel, ``Data-driven techniques for type error diagnosis,'' Ph.D. dissertation, University of California, San Diego, {USA}, 2017. [Online]. Available: \url{http://www.escholarship.org/uc/item/59s4h4pv}
\BIBentrySTDinterwordspacing

\bibitem{difftastic}
\BIBentryALTinterwordspacing
W.~Hughes, ``Difftastic,'' 2021. [Online]. Available: \url{https://github.com/wilfred/difftastic}
\BIBentrySTDinterwordspacing

\bibitem{Typpete}
M.~Hassan, C.~Urban, M.~Eilers, and P.~Müller, ``{MaxSMT-Based Type Inference for Python 3},'' \emph{Computer Aided Verification: 30th International Conference}, pp. 12--19, Jul. 2018.

\bibitem{UnsatCores}
J.~Marques-Silva and J.~Planes, ``{Algorithms for Maximum Satisfiability using Unsatisfiable Cores},'' in \emph{2008 Design, Automation and Test in Europe}, Mar. 2008, pp. 408--413.

\bibitem{TowardsGeneralDiagnosis}
D.~Zhang and A.~C. Myers, ``Toward {General} {Diagnosis} of {Static} {Errors},'' in \emph{Proceedings of the 41st {ACM} {SIGPLAN}-{SIGACT} {Symposium} on {Principles} of {Programming} {Languages}}, Jan. 2014, pp. 569--581.

\bibitem{DiagnosingWithClass}
D.~Zhang, A.~C. Myers, D.~Vytiniotis, and S.~Peyton~Jones, ``Diagnosing type errors with class,'' in \emph{Proceedings of the 36th ACM SIGPLAN Conference on Programming Language Design and Implementation}, vol.~50, Jun. 2015, pp. 12--21.

\bibitem{SHErrLoc}
D.~Zhang, A.~C. Myers, D.~Vytiniotis, and S.~Peyton-Jones, ``{SHErrLoc: A Static Holistic Error Locator},'' \emph{ACM Transactions on Programming Languages and Systems}, vol.~39, no.~4, Aug. 2017.

\bibitem{HMInference}
R.~Milner, ``{A Theory of Type Polymorphism in Programming},'' \emph{Journal of Computer and System Sciences}, vol.~17, no.~3, pp. 348--375, Dec. 1978.

\end{thebibliography}

\appendix

\subsection{Polymorphic Types}

OCaml's type system assigns types to all expressions, for example an integer literal like \lstinline!5!
has type \lstinline!int!. Function types are written with an arrow,
for example a fibonacci function might have type \lstinline!int! $\to$ \lstinline!int!.

Consider an identity function, defined with
\begin{center}
\lstinline!let id x = x;;!
\end{center}
What ought to be the type of this function?
If we infer a type like \lstinline!int $\to$ int! (which is certainly sound), we won't be able to use the function
with booleans, or vice versa. If we assign it the type $\alpha \to \alpha$, where $\alpha$ is a
(monomorphic) type variable, we still have a problem: we can use the function at \lstinline!int! \emph{or}
at \lstinline!bool!, but not both. In fact, this function is frequently passed as an argument to higher-order functions,
and therefore it is common to have it used at many different types throughout a program.

The solution taken by ``Hindley-Milner type systems''~\cite{HMInference} allows \emph{polymorphic} types.
We might express the true type of of \lstinline!id! as $\forall \alpha. \alpha\to\alpha$.
Quantifying over the type variables in a type is called \emph{generalization}.
Whenever the variable \lstinline!id! is referred by the program, a new monomorphic type variable will
be created to represent $\alpha$ for that specific instance, a process called \emph{instantiation}.
Only values bound with a \lstinline!let! binding are generalized -- notably, lambda abstractions are
\emph{not} generalized (unless they are later bound by a \lstinline!let!).

Polymorphic types are a major challenge for type error localization~\cite{Wies,FindingTheSource},
in large part because the generalization and instantiation processes make it difficult to tie a
type mismatch from \emph{outside} of the definition of a \lstinline!let! binding back to a source
in the body of the binding.

\subsection{Classical Type Inference}\label{app:background}

The goal of type inference is to assign a type to every (sub)expression in the program,
thereby ensuring that the program is type-safe, but without requiring any annotations from the programmer.

The classical type inference algorithm described in~\cite{HMInference} proceeds via structural recursion on the program AST.
Each node of the AST corresponds to a (sub)expression of the program.
We use the kind of each subexpression to infer the ``shape'' of its type --
lambda abstractions must have a function type, boolean literals must have the \lstinline!bool! type, etc.
Any unknown information in the inferred shape, such as the input and output types of a function type,
are filled with (monomorphic) type variables. When these type variables correspond to the type of
a named program variable, this relationship is stored in a \emph{context}.

As we recurse through the AST, we may discover relationships between some of the inferred shapes.
For example, when a lambda abstraction is applied to an expression $e$, we learn that the abstraction's
input type must match the type of $e$. We use this information to refine the type variables in both
types through a process called \emph{unification}. Unification ``solves for'' some or all of the type
variables in both types.

A second approach to refining types is to store all of the discovered relationships as
\emph{typing constraints}~\cite{ConstraintForm}. These constraints can be generated for the whole program,
and then later fed into a constraint solver all at once.
We must use such a constraint-based algorithm; see Section \ref{sec:mes} for why.

\end{document}